\documentclass[aps,prd,secnumarabic,nobibnotes,twocolumn,superscriptaddress]{revtex4-1}
\usepackage{amsfonts}
\usepackage{mathrsfs}
\usepackage{amsmath}% needed for subequations
\usepackage{color}
\usepackage{natbib}
\usepackage{graphicx}
\usepackage{bm}% bold maths
\usepackage{amssymb}
\usepackage{xspace}
\usepackage{epstopdf}
\usepackage{dcolumn}
\usepackage{multirow}
\usepackage[colorlinks=true, letterpaper=true, pdfstartview=FitV, linkcolor=blue, citecolor=blue, urlcolor=blue]{hyperref}
\usepackage{wrapfig}

\makeatletter

\newcommand{\Rmnum}[1]{\expandafter\@slowromancap\romannumeral #1@}
\makeatother

\begin{document}
\title{Three-nodal surface phonons in solid-state materials: Theory and material realization}

\author{Chengwu Xie}\thanks{C. X. and H. Y. contributed equally to this manuscript.}
\address{School of Physical Science and Technology, Southwest University, Chongqing 400715, China;}

\author{Hongkuan Yuan}\thanks{C. X. and H. Y. contributed equally to this manuscript.}
\address{School of Physical Science and Technology, Southwest University, Chongqing 400715, China;}

\author{Ying Liu}\email{ying liu@hebut.edu.cn (Y. L.);}
\address{School of Materials Science and Engineering, Hebei University of Technology, Tianjin 300130, China; }

\author{Xiaotian Wang}\email{xiaotianwang@swu.edu.cn (X. W.);}
\address{School of Physical Science and Technology, Southwest University, Chongqing 400715, China;}

\author{Gang Zhang}\email{zhangg@ihpc.a-star.edu.sg (G. Z.);}
\address{Institute of High Performance Computing, Agency for Science, Technology and Research (A*STAR), 138632, Singapore}

\begin{abstract}
This year, Liu \textit{et al}. [Phys. Rev. B \textbf{104}, L041405 (2021)] proposed a new class of topological phonons (TPs; i.e., one-nodal surface (NS) phonons), which provides an effective route for realizing one-NSs in phonon systems. In this work, based on first-principles calculations and symmetry analysis, we extended the types of NS phonons from one- to three-NS phonons. The existence of three-NS phonons (with NS states on the $k_{i}$ = $\pi$ ($i$ = $x$, $y$, $z$) planes in the three-dimensional Brillouin zone (BZ)) is enforced by the combination of two-fold screw symmetry and time reversal symmetry. We screened all 230 space groups (SGs) and found nine candidate groups (with the SG numbers (Nos.) 19, 61, 62, 92, 96, 198, 205, 212, and 213) hosting three-NS phonons. Interestingly, with the help of first-principles calculations, we identified $P2_{1}$2$_{1}$2$_{1}$-type YCuS$_{2}$ (SG No. 19), $Pbca$-type NiAs$_{2}$ (SG No. 61), $Pnma$-type SrZrO$_{2}$ (SG No. 62), $P4_{1}$2$_{1}$2-type LiAlO$_{2}$ (SG No. 92), $P4_{3}$2$_{1}$2-type ZnP$_{2}$ (SG No. 96), $P2_{1}$3-type NiSbSe (SG No. 198), $Pa\bar{3}$-type As$_{2}$Pt (SG No. 205), $P4_{3}$32-type BaSi$_{2}$ (SG No. 212), and $P4_{1}$32-type CsBe$_{2}$F$_{5}$ (SG No. 213) as realistic materials hosting three-NS phonons. The results of our presented study enrich the class of NS states in phonon systems and provide concrete guidance for searching for three-NS phonons and singular Weyl point phonons in realistic materials.
\end{abstract}
\maketitle

%%%%%%% Main text %%%%%%%%%%%%%%%%%%%%%
\section{Introduction}
Topological quantum states of matter~\cite{add1,add2,add3} are an important topic in the field of modern condensed-matter physics. Over the past 15 years, we have witnessed the emergence of many types of topological electronic materials, such as topological insulators~\cite{add4,add5,add6,add7}, topological crystalline insulators~\cite{add8,add9,add10}, topological Kondo insulators~\cite{add11,add12,add13}, higher-order topological insulators~\cite{add14,add15,add16,add17}, topological semimetals~\cite{add18,add19,add20}, and higher-order topological semimetals~\cite{add21,add22,add23,add24}. In particular, the types and numbers of topological semimetals~\cite{add20} are rapidly increasing. In contrast to Dirac, Weyl, and Majorana fermions, which are allowed in high-energy physics, the types of quasiparticles in topological semimetals~\cite{add25} are more diverse owing to fewer constraints imposed by the space group (SG) symmetries of the crystal. Based on the dimensionality of the band-crossings in the momentum space, the topological semimetals can be classified into nodal point~\cite{add26,add27,add28,add29,add30}, nodal line~\cite{add31,add32,add33,add34,add35}, and nodal surface (NS)~\cite{add36,add37,add38,add39,add40} semimetals with zero-, one-, and two-dimensional band-crossings, respectively.

\begin{figure}[t]
\includegraphics[width=8cm]{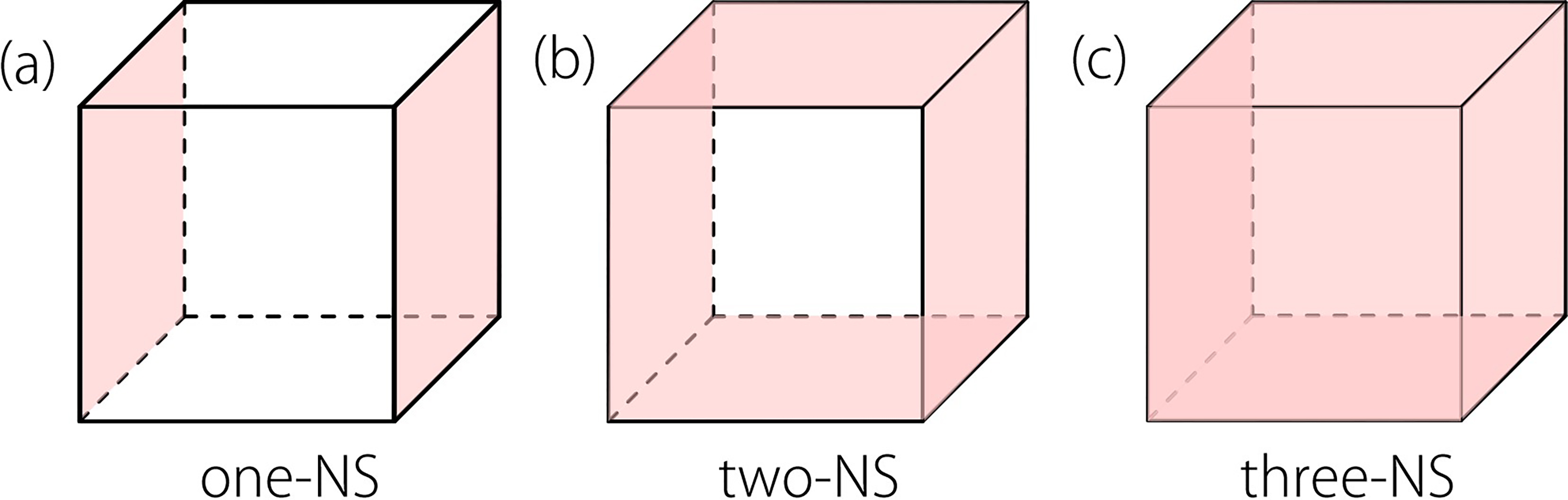}
\caption{Schematic diagrams of (a) one-NS, (b) two-NS, (c) and three-NS phonons, respectively.
\label{fig1}}
\end{figure}

Three-dimensional topological semimetals with two-dimensional band-crossings can host NS states in the Brillouin zone (BZ). Each point on the NS should be a two-fold degenerate point with linear band dispersion along the surface normal direction. Researchers hope that NS semimetals exhibit exotic physical properties, such as stronger quantum oscillations and peculiar plasmon excitations. Wu \textit{et al}.~\cite{add36} summarized an essential NS state dictated by nonsymmorphic symmetry without spin-orbit coupling (SOC). The existence of a series of NS semimetals in realistic electronic systems has been predicted, including BaVS$_{3}$~\cite{add40}, ZrSiS~\cite{add41,add42}, K$_{6}$YO$_{4}$~\cite{add36}, FeB$_{4}$~\cite{add43}, Ti$_{3}$Al~\cite{add37}, and X(MoS)$_{3}$ (X = K, Rb, and Cs)~\cite{add44}. However, in general, SOC in electronic materials cannot be ignored; thus, the proposed two-dimensional nonsymmorphic symmetry-enforced NS states in electronic systems will usually be destroyed or reduced to one-dimensional nodal lines when SOC is considered~\cite{add42}. Moreover, the NS states in some materials are far from the Fermi level and exhibit large energy variations, which hinder their experimental detection.
\\
 \indent{The proposed topological phonons (TPs)~\cite{add45,add46} have renewed the interest in topological quantum states; TPs are a basic kind of boson-type quasiparticles; they are not affected by the Pauli exclusion principle and SOC. Therefore, TPs can normally be observed in spinless phononic systems in all frequency ranges. In addition to the proposed nodal point phonons~\cite{add47,add48,add49,add50,add51,add52,add53,add54,add55} and nodal line phonons~\cite{add56,add57,add58,add59,add60,add61,add62,add63,add64}, one-NS phonons~\cite{add65} have been presented by Liu \textit{et al}. based on symmetry analysis and first-principles calculations. The researchers provided a complete list of the one-NS phonons in the 230 SGs and discovered that RbTeAu family materials with SG number (No.) 51 may contain one-NS states (on the $k_{x}$ = $\pi$ plane). The occurrence of one-NS states is ensured by screw rotation symmetry along the $i$ axis ($i$ = $x$, $y$, or $z$) and time-reversal symmetry $\mathcal{T}$. Fig.~\ref{fig1}(a) presents a schematic diagram of one-NS phonons. Moreover, two more types of NS phonons should exist: two- and three-NS phonons, as illustrated in Fig.~\ref{fig1}(b) and Fig.~\ref{fig1}(c), respectively.}
\\
\indent{In this study, we extended the class of NS phonons from one- to three-NS phonons. For the three-NS phonons, the NS states are localized on the $k_{i}$ = $\pi$ ($i$ = $x$, $y$, $z$) planes in the three-dimensional BZ. We screened all 230 SGs; the SGs with Nos. 19, 61, 62, 92, 96, 198, 205, 212, and 213 are candidate groups that can obtain three-NS phonons. Because the three-NS phonons in these SGs are symmetry-enforced, one can easily achieve three-NS phonons in realistic materials with the previously presented SGs. For example, in this work, we identified $P2_{1}$2$_{1}$2$_{1}$-type YCuS$_{2}$ (SG No. 19), $Pbca$-type NiAs$_{2}$ (SG No. 61), $Pnma$-type SrZrO$_{2}$ (SG No. 62), $P4_{1}$2$_{1}$2-type LiAlO$_{2}$ (SG No. 92), $P4_{3}$2$_{1}$2-type ZnP$_{2}$ (SG No. 96), $P2_{1}$3-type NiSbSe (SG No. 198), $Pa\bar{3}$-type As$_{2}$Pt (SG No. 205), $P4_{3}$32-type BaSi$_{2}$ (SG No. 212), and $P4_{1}$32-type CsBe$_{2}$F$_{5}$ (SG No. 213) as realistic materials that can host three-NS phonons.}

\section{Symmetry Analysis of Three-NS Phonons.}

In this part, we searched all essential NSs, which are only dictated by symmetries, in spinless systems~\cite{add36}. Such an NS is protected by the combination of time-reversal symmetry ($\mathcal{T}$) and two-fold screw rotation symmetry ($S_{2i}$).

Without loss of generalization, we take two-fold screw rotation along the z-direction as an example: $S_{2z}$:$(x,y,z) \to (-x,-y,z+\frac{1}{2})$ with a half translation in the lattice constant along its rotation axis. It also affects the momentum space: $S_{2z}$:$(k_{x},k_{y},k_{z}) \to (-k_{x},-k_{y},k_{z})$, thereby only preserving the momentum along $k_{z}$. Without SOC, $S_{2z}^{2}$=$T_{100}$=$e^{-ik_{z}}$, where $T_{100}$ is the translation along the z-direction. For time-reversal symmetry, in spinless systems, $\mathcal{T}^{2}$ = 1, which is antiunitary and inverses the momentum \textbf{\textit{k}}. Consequently, their combination $\mathcal{T}S_{2z}$ is also antiunitary. Remarkably, on planes where $k_{z}$ = $\pm\pi$, $(\mathcal{T}S_{2z})^{2}$=$e^{-ik_{z}}|_{k_{z}=\pm\pi}$ = $-1$, which suggests Kramer-like degeneracy on these planes. Thereby, it leads to Kramer-like degeneracy. Hence, the phonon bands on the $k_{i}$ = $\pi$ ($i$ = $x$, $y$, $z$) planes must become two-fold degenerate, thereby forming three-NS phonons. Furthermore, the presence of three two-fold rotation symmetries (i.e., $S_{2x}$, $S_{2y}$, and $S_{2z}$) leads to three NSs on the planes $k_{i}$ = $\pm\pi$ ($i$ = $x$, $y$, $z$). In this study, we proposed all the three-NS phonons by searching all 230 SGs in phonon systems. According to the results, the SGs with Nos. 19, 61, 62, 92, 96, 198, 205, 212, and 213 (see Table~\ref{table1}) can host three-NS phonons.

\section{Computational Details}

First-principles calculations based on density functional theory were performed to study the ground states of $P2_{1}$2$_{1}$2$_{1}$-type YCuS$_{2}$, $Pbca$-type NiAs$_{2}$, $Pnma$-type SrZrO$_{2}$, $P4_{1}$2$_{1}$2-type LiAlO$_{2}$, $P4_{3}$2$_{1}$2-type ZnP$_{2}$, $P2_{1}$3-type NiSbSe, $Pa\bar{3}$-type As$_{2}$Pt, $P4_{3}$32-type BaSi$_{2}$, and $P4_{1}$32-type CsBe$_{2}$F$_{5}$ materials, as implemented in the Vienna Ab Initio Simulation Package. The projector augmented wave method and generalized gradient approximation~\cite{add66} with Perdew--Burke--Ernzerhof functions were used for the ionic potential and exchange-correlation interaction. In addition, a plane wave cutoff energy of 500 eV was used for the structural relaxation. The following $k$-mesh samples were used for YCuS$_{2}$, NiAs$_{2}$, SrZrO$_{2}$, LiAlO$_{2}$, ZnP$_{2}$, NiSbSe, As$_{2}$Pt, BaSi$_{2}$, and CsBe$_{2}$F$_{5}$: $9\times7\times5$, $5\times5\times3$, $5\times5\times5$, $7\times7\times7$, $7\times7\times3$, $7\times7\times7$, $7\times7\times7$, $9\times9\times9$, and $5\times5\times5$, respectively. All these materials are experimentally synthesized materials. The phononic dispersions of the $2\times2\times1$ YCuS$_{2}$, $2\times2\times1$ NiAs$_{2}$, $2\times1\times2$ SrZrO$_{2}$, $2\times2\times2$ LiAlO$_{2}$, $2\times2\times1$ ZnP$_{2}$, $2\times2\times2$ NiSbSe, $2\times2\times2$ As$_{2}$Pt, $2\times2\times2$ BaSi$_{2}$, and $1\times1\times1$ CsBe$_{2}$F$_{5}$ cells were examined with density functional perturbation theory and PHONOPY codes~\cite{add67}.

\begin{table*}[t]
\centering
\small
\renewcommand\arraystretch{1.3}
  \caption{A complete list of three-NS phonons in 230 SGs. The first and second columns present the SG numbers and SG symbols, the third column lists the three-NSs along the symmetry paths, and the fourth column presents the corresponding realistic materials.}
  \label{table1}
  \begin{tabular*}{0.85\textwidth}{@{\extracolsep{\fill}}cccc}
  \hline
  \hline
  Space group & Space group & Three-nodal & Realistic  \\
    numbers   & symbols     &  surfaces & materials \\
    \hline
   19 & $P2_{1}$2$_{1}$2$_{1}$ & NS$_{TYS}$, NS$_{SXU}$, and NS$_{UZT}$ & YCuS$_{2}$  \\
   61 & $Pbca$                 & NS$_{TYS}$, NS$_{SXU}$, and NS$_{UZT}$ & NiAs$_{2}$  \\
   62 & $Pnma$                 & NS$_{TYS}$, NS$_{SXU}$, and NS$_{UZT}$ & SrZrO$_{2}$  \\
   92 & $P4_{1}$2$_{1}$2       & NS$_{ZRA}$, and NS$_{MXR}$             & LiAlO$_{2}$  \\
   96 & $P4_{3}$2$_{1}$2       & NS$_{ZRA}$, and NS$_{MXR}$             & ZnP$_{2}$  \\
  198 & $P2_{1}$3              & NS$_{RXM}$                             & NiSbSe  \\
  205 & $Pa\bar{3}$            & NS$_{RXM}$                             & As$_{2}$Pt  \\
  212 & $P4_{3}$32             & NS$_{RXM}$                             & BaSi$_{2}$  \\
  213 & $P4_{1}$32             & NS$_{RXM}$                             & CsBe$_{2}$F$_{5}$  \\
    \hline
    \hline
  \end{tabular*}
\end{table*}

\section{Materials with Three-NS phonons}

\begin{figure}[t]
\includegraphics[width=5cm]{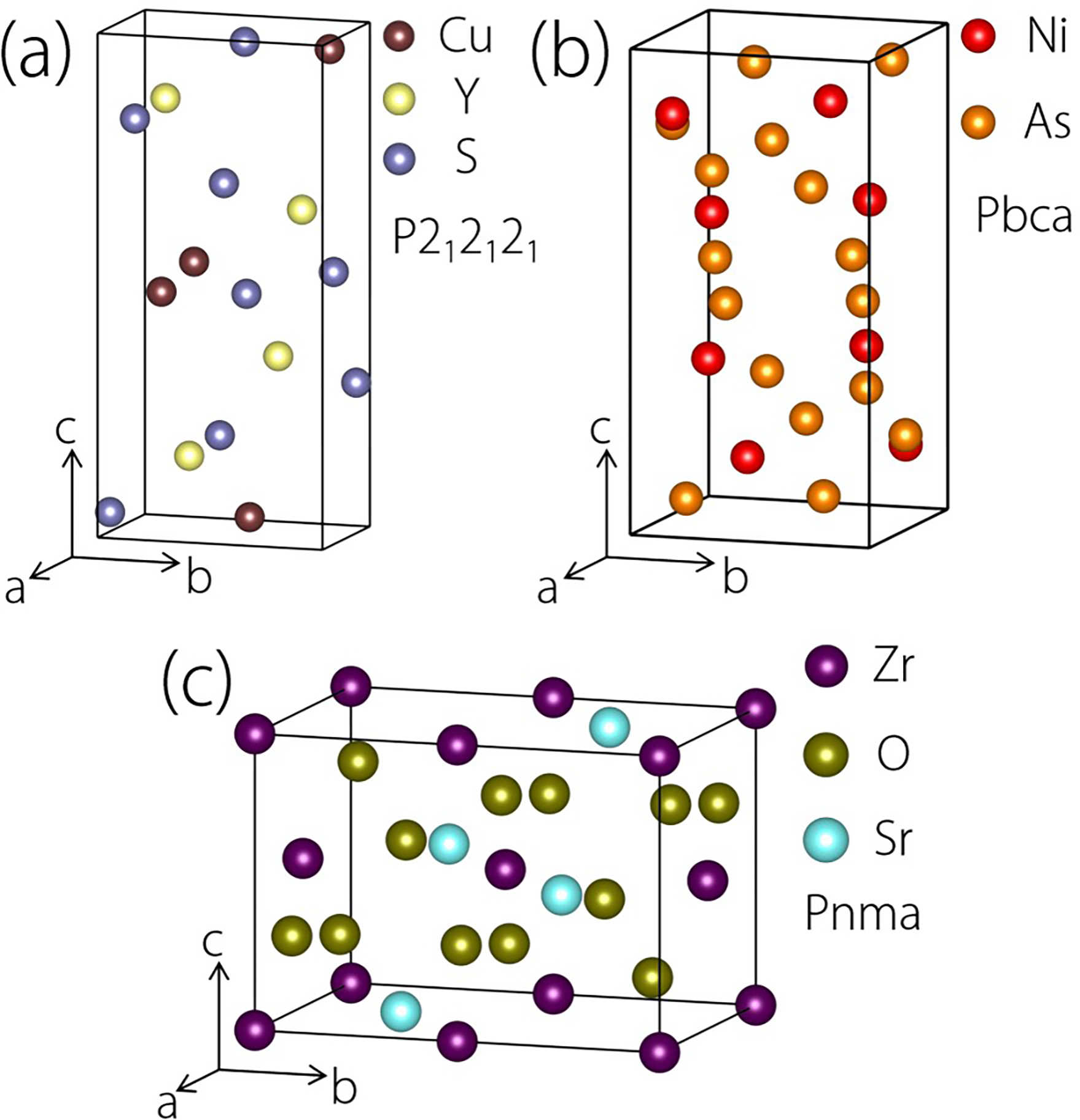}
\caption{Crystal structures of $P2_{1}$2$_{1}$2$_{1}$-type YCuS$_{2}$ (SG No. 19), $Pbca$-type NiAs$_{2}$ (SG No. 61), and $Pnma$-type SrZrO$_{2}$ (SG No. 62), respectively.
\label{fig2}}
\end{figure}

For phonon systems with SG Nos. 19, 61, 62, the three-NSs (i.e., NS$_{TYS}$, NS$_{SXU}$, and NS$_{UZT}$) appear on the planes $k_{y}$ = $\pi$, $k_{x}$ = $\pi$, and $k_{z}$ = $\pi$, respectively. Some realistic materials were selected as examples to demonstrate that they host three-NSs in their phonon dispersions: $P2_{1}$2$_{1}$2$_{1}$-type YCuS$_{2}$ (SG No. 19) can be prepared~\cite{add68} by fusing the high-purity elements in evacuated quartz ampoules; Murray and Heyding~\cite{add69} prepared $Pbca$-type NiAs$_{2}$ (SG No. 61) by heating the elements in sealed evacuated Vycor tubes; $Pnma$-type SrZrO$_{2}$ powders (SG No. 62) were prepared with the polymeric precursor method by Cavalcante \textit{et al}.~\cite{add70}. The crystal structures of these three materials are shown in Fig.~\ref{fig2}. They are completely relaxed; their theoretically determined lattice constants and the previously published experimentally determined data are listed in Table~\ref{table2}.

\begin{table}[t]
\centering
\small
\renewcommand\arraystretch{1.3}
  \caption{Theoretically and experimentally determined lattice constants of YCuS$_{2}$, NiAs$_{2}$, and SrZrO$_{2}$.}
  \label{table2}
  \begin{tabular*}{0.50\textwidth}{@{\extracolsep{\fill}}ccc}
  \hline
  \hline
    Materials & Theoretical lattice & Experimental lattice    \\
            & constants           & constants~\cite{add68,add69,add70}  \\
    \hline
    YCuS$_{2}$  & a = 3.96 \AA, b = 6.26 \AA,  & a = 3.97 \AA, b = 6.27 \AA,  \\
              & c = 13.48 \AA                & c = 13.38 \AA                \\
    NiAs$_{2}$  & a = 5.80 \AA, b = 5.89 \AA,  & a = 5.77 \AA, b = 5.83 \AA,  \\
              & c = 11.50 \AA                & c = 11.41 \AA                \\
    SrZrO$_{2}$ & a = 5.91 \AA, b = 8.29 \AA,  & a = 5.81 \AA, b = 8.19 \AA,  \\
              & c = 5.84 \AA                 & c = 5.79 \AA                 \\
    \hline
    \hline
  \end{tabular*}
\end{table}

\begin{figure}[t]
\includegraphics[width=8cm]{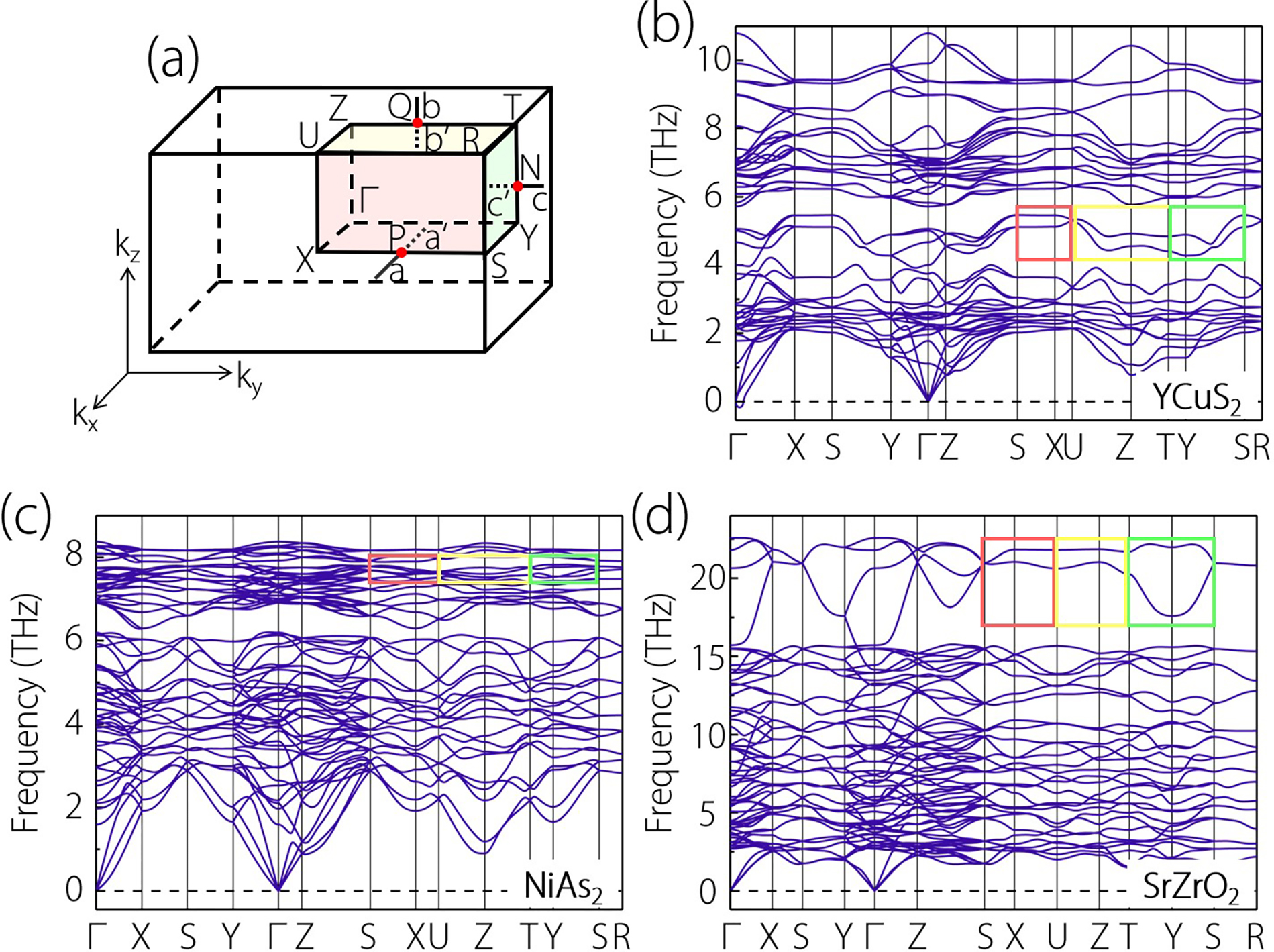}
\caption{(a) Three-dimensional BZ and symmetry points. Three-NS states (red, green, and yellow) are localized on the $k_{i}$ = $\pi$ ($i$ = $x$, $y$, $z$) planes in three-dimensional BZ. (b)--(d) Calculated phonon dispersions of YCuS$_{2}$, NiAs$_{2}$, and SrZrO$_{2}$, respectively. Three NS regions (i.e., NS$_{TYS}$, NS$_{SXU}$, and NS$_{UZT}$) are highlighted in green, red, and yellow, respectively.
\label{fig3}}
\end{figure}

\begin{figure}[t]
\includegraphics[width=7cm]{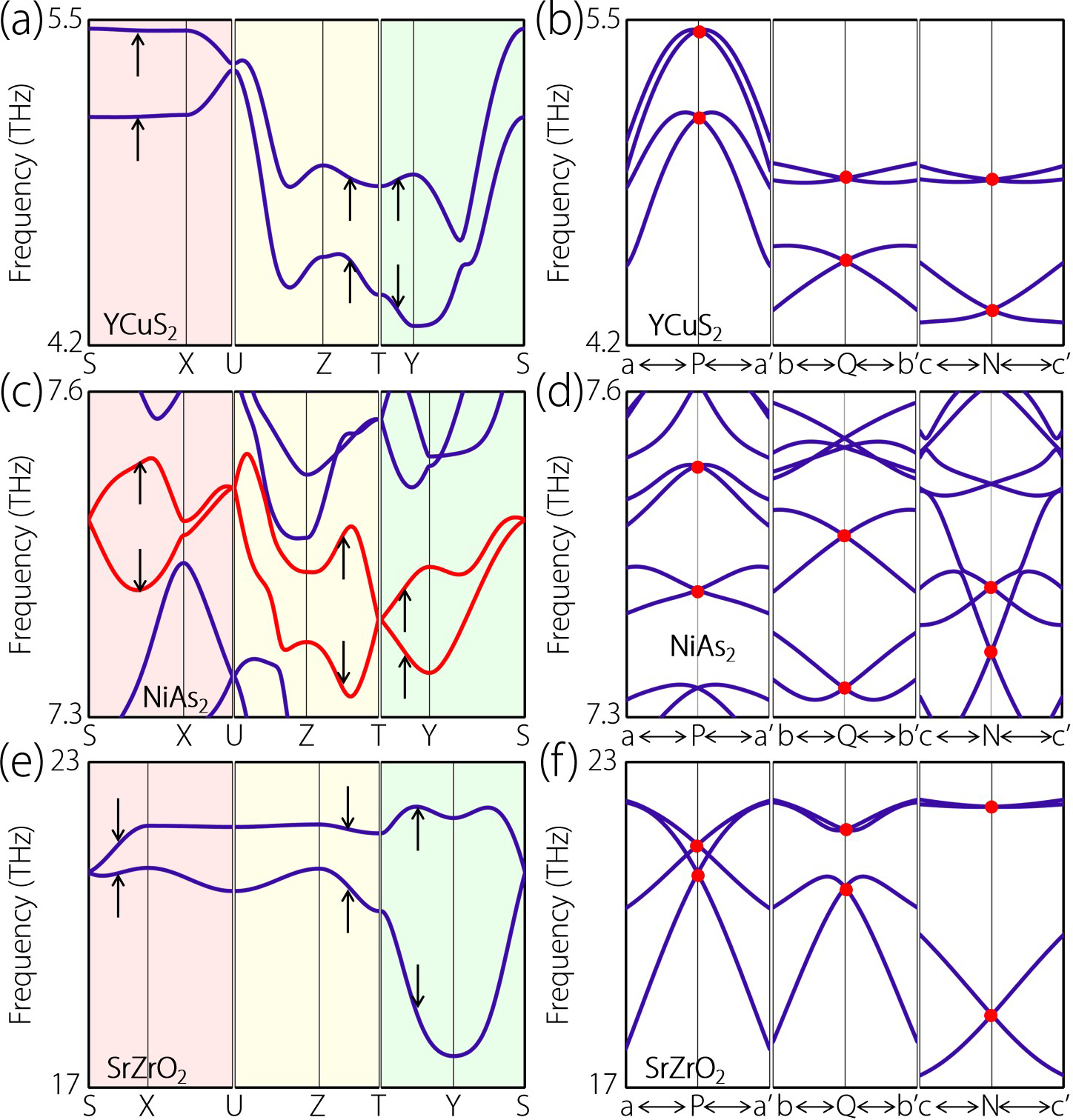}
\caption{(a), (c), and (e) Enlarged phonon dispersions of three regions (see Fig.~\ref{fig3}(b)--(d)) of YCuS$_{2}$, NiAs$_{2}$, and SrZrO$_{2}$, respectively. (b), (d), and (f) phonon dispersions along $a$ (0.70, 0.25, 0.00) -$P$ (0.50, 0.25, 0.00) -$a^{\prime}$ (0.30, 0.25, 0.00), $b$ (0.00, 0.25, 0.7) -$Q$ (0.00, 0.25, 0.5) -$b^{\prime}$ (0.00, 0.25, 0.3), and $c$ (0.00, 0.7, 0.25) -$N$ (0.00, 0.5, 0.25) -$c^{\prime}$ (0.00, 0.3, 0.25), respectively. All points at $P$, $Q$, and $N$ points are two-fold degenerate with linear phonon band dispersions.
\label{fig4}}
\end{figure}

The phonon dispersions of YCuS$_{2}$, NiAs$_{2}$, and SrZrO$_{2}$ along the $\Gamma - X - S - Y - \Gamma - Z - S - X - U - Z - T - Y - S -R$ paths (see Fig.~\ref{fig3}(a)) are shown in Fig.~\ref{fig3}(b)--(d), respectively. Three regions (highlighted in red, yellow, and green) are of interests in this study. The enlarged figures of the phonon dispersions of YCuS$_{2}$, NiAs$_{2}$, and SrZrO$_{2}$ in the three regions are shown in Fig.~\ref{fig4}(a), (c) and (e), respectively. All the phonon bands along the $S$-$X$-$U$, $U$-$Z$-$T$, and $T$-$Y$-$S$ planes have two-fold degeneracy. To explain this in more detail, some symmetry lines (i.e., $a$-$P$-$a^{\prime}$, $b$-$Q$-$b^{\prime}$, and $c$-$N$-$c^{\prime}$; see Fig.~\ref{fig3}(a)) are selected, they are perpendicular to the $S$-$X$, $T$-$Z$, and $Y$-$T$ symmetry lines, respectively. Subsequently, we calculate the phonon dispersions along the $a$-$P$-$a^{\prime}$, $b$-$Q$-$b^{\prime}$, and $c$-$N$-$c^{\prime}$ paths; the results are presented in Fig.~\ref{fig4}(b), (d), (f), respectively. Evidently, the points (highlighted by red circles) at the $P$, $Q$, and $N$ symmetry points are two-fold degenerate and have linear band dispersions. In addition, two-fold Kramer-like degeneracy occurs at every point on the $S$-$X$-$U$, $U$-$Z$-$T$, and $T$-$Y$-$S$ planes, thereby forming three-NSs on the $k_{i}$ = $\pi$ ($i$ = $x$, $y$, $z$) planes. These density functional theory results agree well with the argument (Section II) that the antiunitary symmetry $\mathcal{T}S_{2i}$ ensures the existence of three-NS phonons on the $k_{i}$ = $\pm\pi$ ($i$ = $x$, $y$, $z$) planes.

\begin{figure}[t]
\includegraphics[width=7.5cm]{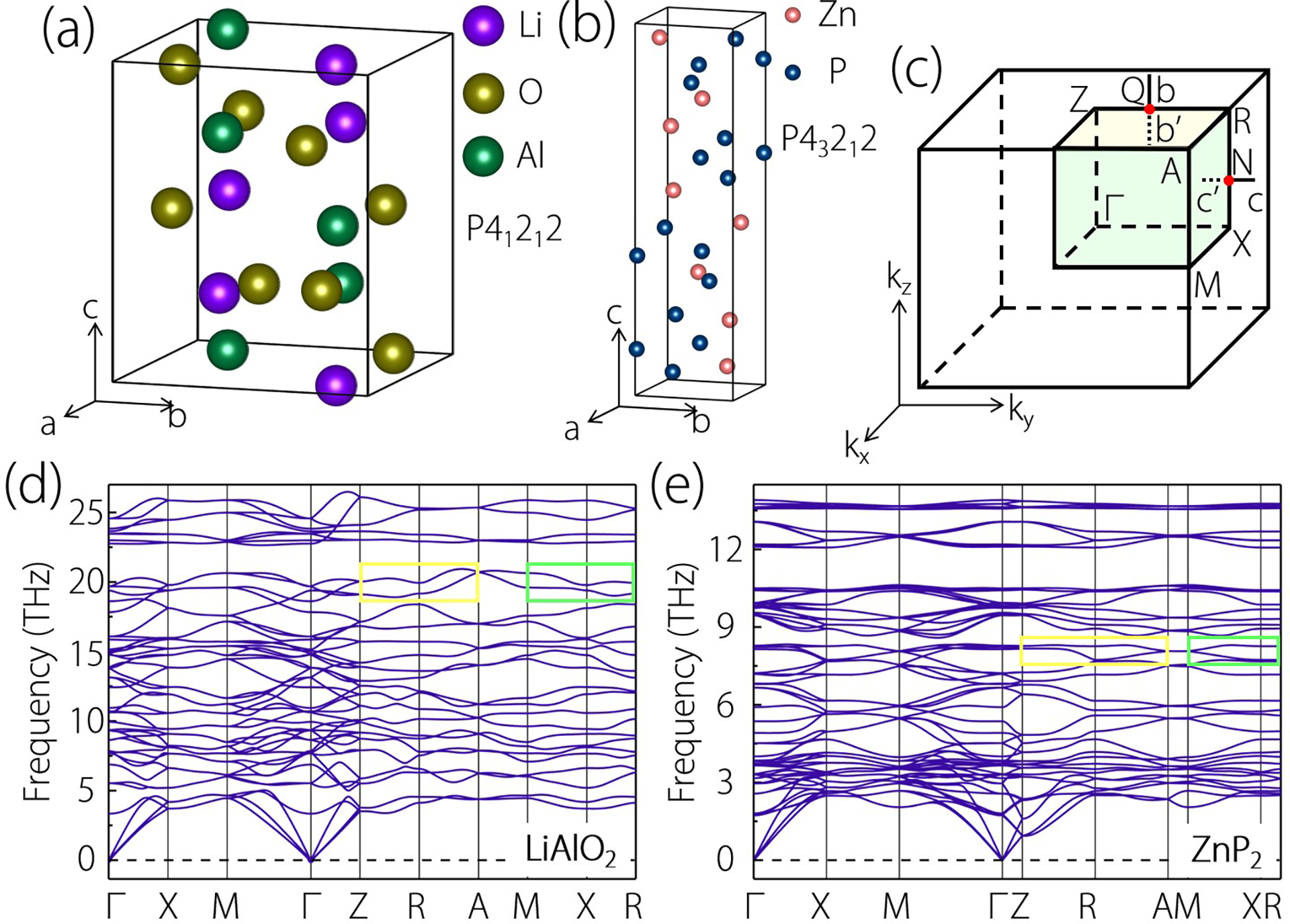}
\caption{(a) and (b) Crystal structures of $P4_{1}$2$_{1}$2-type LiAlO$_{2}$ (SG No. 92) and $P4_{3}$2$_{1}$2-type ZnP$_{2}$ (SG No. 96), respectively; (c) Three-dimensional BZ and symmetry points. Three-NS states (highlighted in yellow and green) are localized on $k_{i}$ = $\pi$ ($i$ = $x$, $y$, $z$) planes in three-dimensional BZ. (b)--(d) Calculated phonon dispersions of LiAlO$_{2}$ and ZnP$_{2}$, respectively. NS regions (i.e., NS$_{MXR}$, and NS$_{ZRA}$) are highlighted in green and yellow, respectively.
\label{fig5}}
\end{figure}

\begin{figure}[t]
\includegraphics[width=7.5cm]{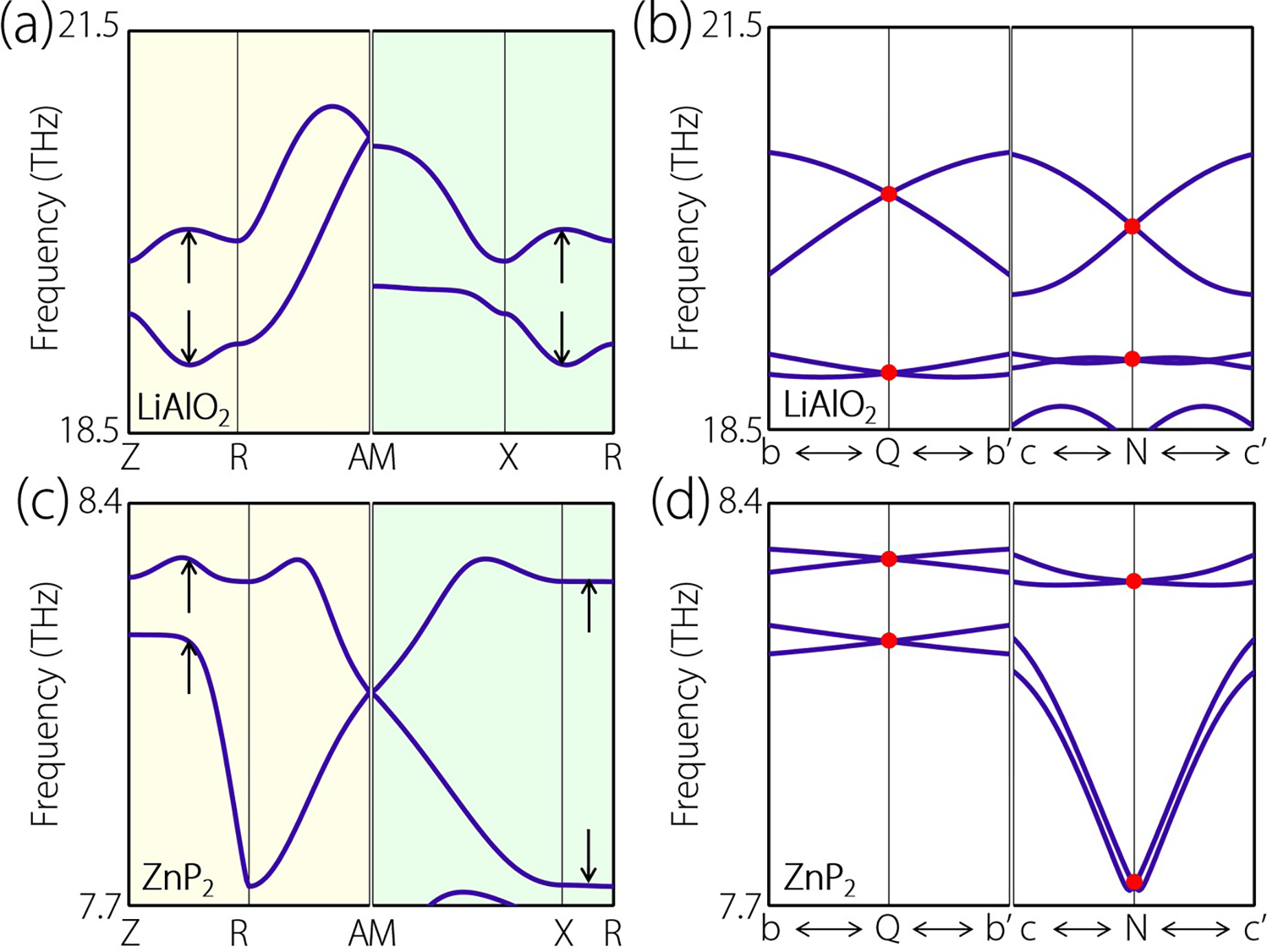}
\caption{(a), (c) Enlarged phonon dispersions of two regions (see Fig.~\ref{fig5}(d) and (e)) of LiAlO$_{2}$ and ZnP$_{2}$, respectively. (b), (d) Phonon dispersions along $b$ (0.00, 0.25, 0.7) -$N$ (0.00, 0.25, 0.5) -$b^{\prime}$ (0.00, 0.25, 0.3), and $c$ (0.00, 0.7, 0.25) -$Q$ (0.00, 0.5, 0.25) -$c^{\prime}$ (0.00, 0.3, 0.25), respectively. All the points at the $Q$ and $N$ points are two-fold degenerate points with linear phonon band dispersions.
\label{fig6}}
\end{figure}

In the next step, some realistic materials with SG Nos. 92 and 96 and three-NS phonons are presented. The first example is $P4_{1}$2$_{1}$2-type LiAlO$_{2}$ (SG No. 92); Remeika and Ballman~\cite{add71} prepared these single crystals from a flux. The second example is $P4_{3}$2$_{1}$2-type ZnP$_{2}$ (SG No. 96). Researchers~\cite{add72} have reported that ZnP$_{2}$ crystals can exist in an enantiomorphic form with SG $P4_{3}$2$_{1}$2 = $D_{4}^{8}$. The theoretically determined and previously published experimental lattice constants are shown in Table~\ref{table3}. The crystal structures of the two materials are shown in Fig.~\ref{fig5}(a) and (b). The phonon dispersions of the two materials along the $\Gamma - X - M - \Gamma - Z - R - A - M - X - R$ paths (see Fig.~\ref{fig5}(c)) are presented in Fig.~\ref{fig5}(d) and (e). To examine the three-NS phonons in these two materials, we only focused on two paths: $Z$-$R$-$A$ and $M$-$X$-$R$, which are highlighted in yellow and green in Fig.~\ref{fig5}(d) and (e), respectively. The enlarged phonon dispersions of these two regions for LiAlO$_{2}$ and ZnP$_{2}$ are shown in Fig.~\ref{fig6}(a) and (c), respectively. All the phonon bands along the $Z$-$R$-$A$ and $M$-$X$-$R$ paths are two-fold degenerate. To present examples, we select two symmetry points $N$ and $Q$ on the $k_{y}$ = $\pi$ and $k_{z}$ = $\pi$ planes, respectively. We construct the two paths $b$-$Q$-$b^{\prime}$ and $c$-$N$-$c^{\prime}$, which vertically pass through the $k_{z}$ = $\pi$ and $k_{y}$ = $\pi$ planes, respectively. The obtained phonon dispersions along the $b$-$Q$-$b^{\prime}$ and $c$-$N$-$c^{\prime}$ paths for LiAlO$_{2}$ and ZnP$_{2}$ are shown in Fig.~\ref{fig6}(b) and (d), respectively. There are two two-fold degenerate points at $Q$ and $N$, which represent the NS states on the $k_{z}$ = $\pi$ and $k_{y} = \pi$ planes. Because LiAlO$_{2}$ and ZnP$_{2}$ with SG Nos. 92 and 96 host four-fold screw rotation, $S_{4z}$ = $\{ C_{4z}|00\frac{1}{2} \}$, there should be NSs on the $k_{x}$ = $\pm\pi$ planes.

\begin{table}[t]
\centering
\small
\renewcommand\arraystretch{1.3}
  \caption{Theoretically and experimentally determined lattice constants of LiAlO$_{2}$ and ZnP$_{2}$.}
  \label{table3}
  \begin{tabular*}{0.50\textwidth}{@{\extracolsep{\fill}}ccc}
  \hline
  \hline
    Materials & Theoretical lattice & Experimental lattice    \\
            & constants           & constants~\cite{add71,add72}  \\
    \hline
    LiAlO$_{2}$  & a = b = 5.21 \AA,  & a = b = 5.17 \AA, \\
                 & c = 6.30 \AA       & c = 6.59 \AA      \\
    ZnP$_{2}$    & a = b = 5.06 \AA,  & a = b = 5.10 \AA,  \\
                 & c = 18.53 \AA      & c = 18.62 \AA       \\
    \hline
    \hline
  \end{tabular*}
\end{table}

\begin{figure}[t]
\includegraphics[width=7.5cm]{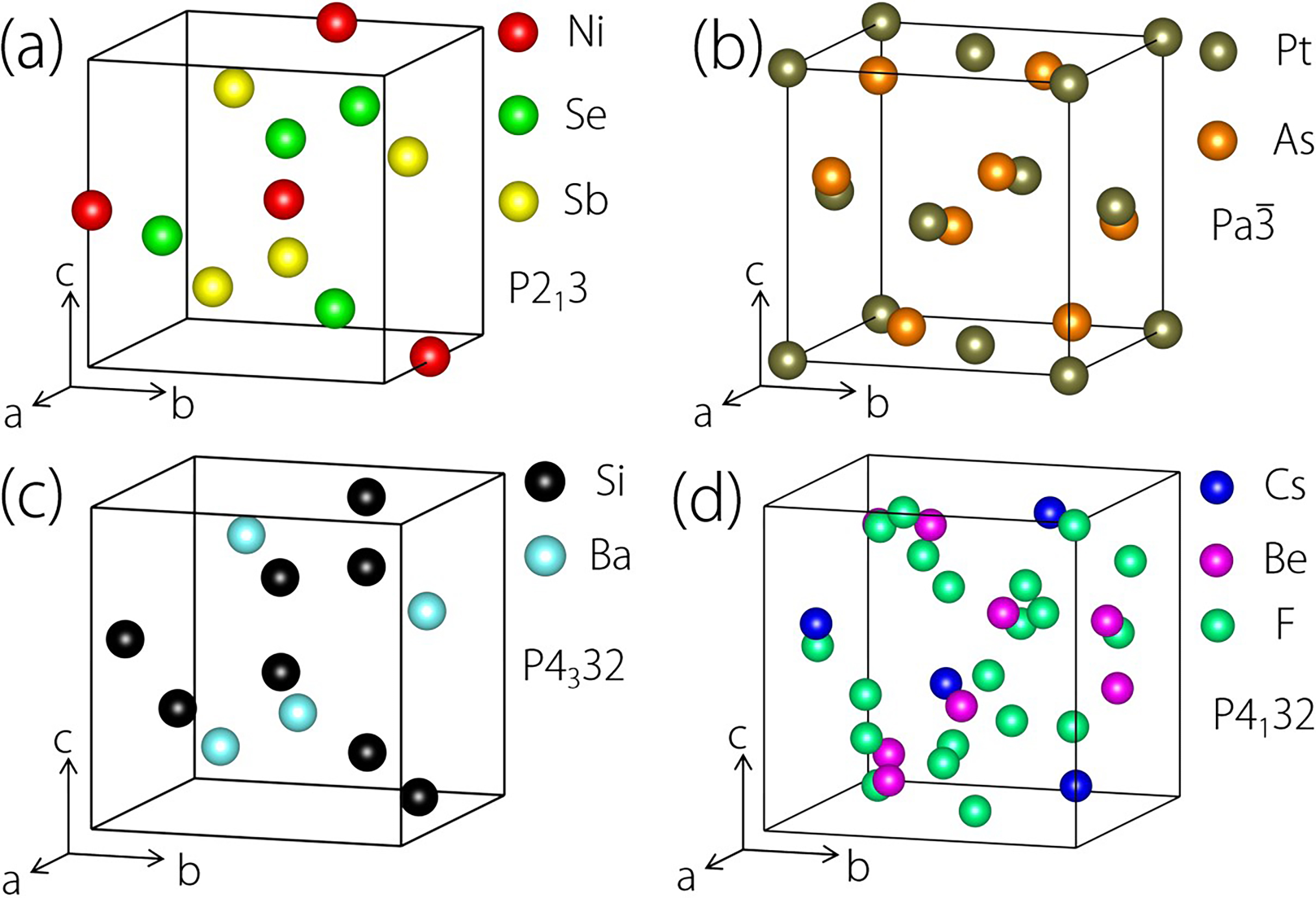}
\caption{(a)--(d) Crystal structures of $P2_{1}$3-type NiSbSe (SG No. 198), $Pa\bar{3}$-type As$_{2}$Pt (SG No. 205), $P4_{3}$32-type BaSi$_{2}$ (SG No. 212), and $P4_{1}$32-type CsBe$_{2}$F$_{5}$ (SG No. 213), respectively.
\label{fig7}}
\end{figure}

Finally, some realistic materials with SG Nos. 198, 205, 212, and 213 are presented, which host three-NS phonons, i.e., NS$_{RXM}$. The first example is $P2_{1}$3-type NiSbSe (SG No. 198). It was prepared by letting powders of binary nickel chalcogenides react with the respective pnictogen component in evacuated sealed silica tubes~\cite{add73}. The second example is $Pa\bar{3}$-type As$_{2}$Pt (SG No. 205). Ramsdell~\cite{add74} produced artificial PtAs$_{2}$, which is identical to natural sperrylite. The third example is $P4_{3}$32-type BaSi$_{2}$ (SG No. 212). This compound is an interesting material~\cite{add75} that can host three types of polymorphs (orthorhombic, trigonal, and cubic crystal classes with $Pnma$, $P\bar{3}m1$, and $P4_{3}32$ SGs) at up to 40 kbar and 1000 $^{\circ}C$. $P4_{3}$32-type BaSi$_{2}$ represents one kind of the high-pressure phases. The fourth example is $P4_{1}$32-type CsBe$_{2}$F$_{5}$ (SG No. 213). Le Fur and Al$\rm{\acute{e}}$onard~\cite{add76} dissolved Cs$_{2}$CO$_{2}$ carbonate in a hydrofluoric solution containing excess BeF$_{2}$. A single CsBe$_{2}$F$_{5}$ crystal can be obtained via evaporation at 55 $^{\circ}C$. The crystal structures of these materials are shown in Fig.~\ref{fig7}. We determined their lattice constants with structural-relaxation calculations (see Fig.~\ref{fig7} and Table~\ref{table4}).

\begin{table}[t]
\centering
\small
\renewcommand\arraystretch{1.3}
  \caption{Theoretically and experimentally determined lattice constants of NiSbSe, As$_{2}$Pt, BaSi$_{2}$, and CsBe$_{2}$F$_{5}$}
  \label{table4}
  \begin{tabular*}{0.50\textwidth}{@{\extracolsep{\fill}}ccc}
  \hline
  \hline
    Materials & Theoretical lattice & Experimental lattice    \\
            & constants           & constants~\cite{add73,add74,add75,add76}  \\
    \hline
    NiSbSe             & a = b = c = 6.13 \AA  & a = b = c = 6.08 \AA \\
    As$_{2}$Pt         & a = b = c = 6.06 \AA  & a = b = c = 5.92 \AA  \\
    BaSi$_{2}$         & a = b = c = 6.77 \AA  & a = b = c = 6.71 \AA  \\
    CsBe$_{2}$F$_{5}$  & a = b = c = 8.06 \AA  & a = b = c = 7.93 \AA  \\
    \hline
    \hline
  \end{tabular*}
\end{table}

\begin{figure}[t]
\includegraphics[width=8.5cm]{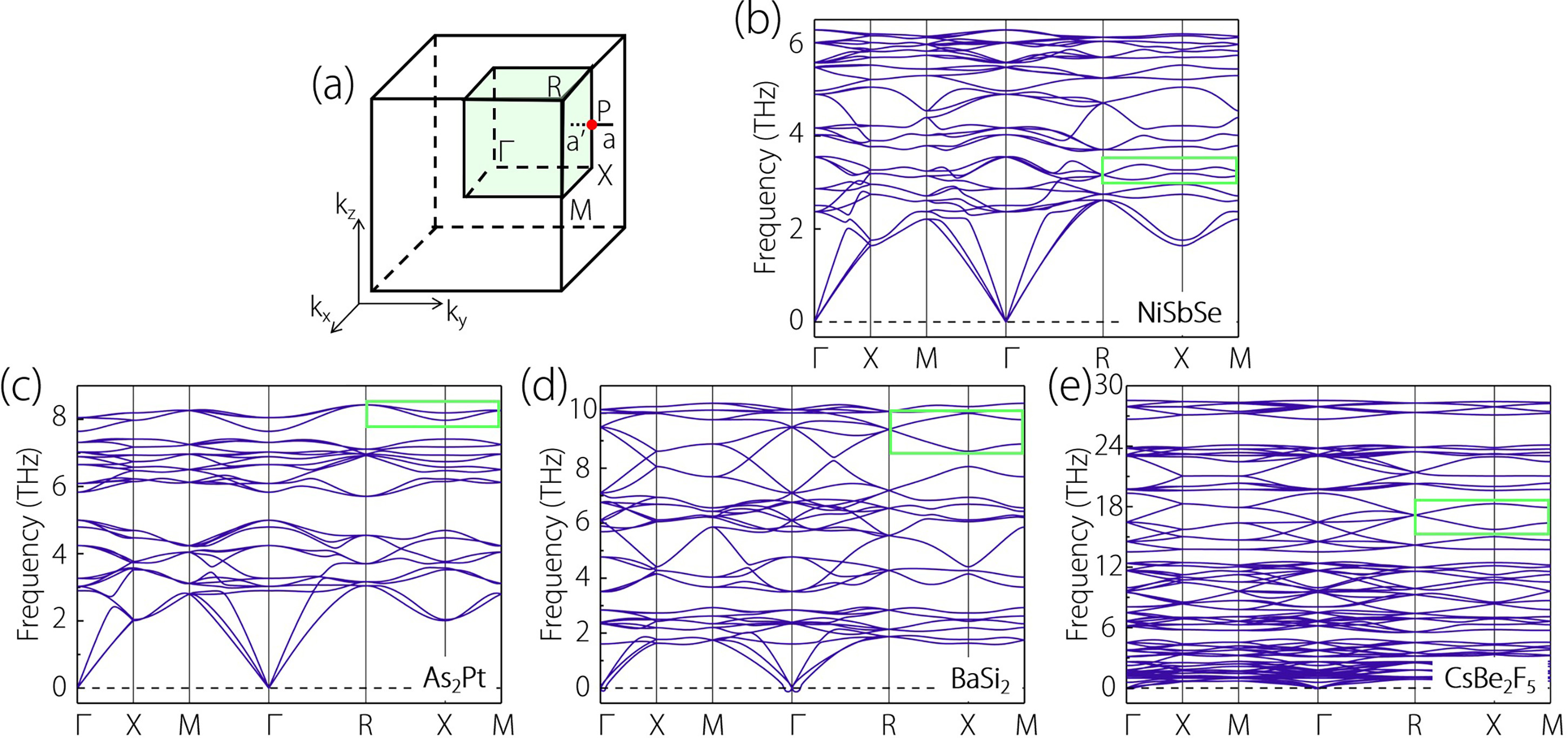}
\caption{(a) Three-dimensional BZ and symmetry points. Three-NS states (green color) are localized on $k_{i}$ = $\pi$ ($i$ = $x$, $y$, $z$)  planes in three-dimensional BZ. (b)--(e) Calculated phonon dispersions of NiSbSe, As$_{2}$Pt, BaSi$_{2}$, and CsBe$_{2}$F$_{5}$ materials, respectively. NS regions (i.e., NS$_{RXM}$) are highlighted in green.
\label{fig8}}
\end{figure}

\begin{figure}[t]
\includegraphics[width=7.5cm]{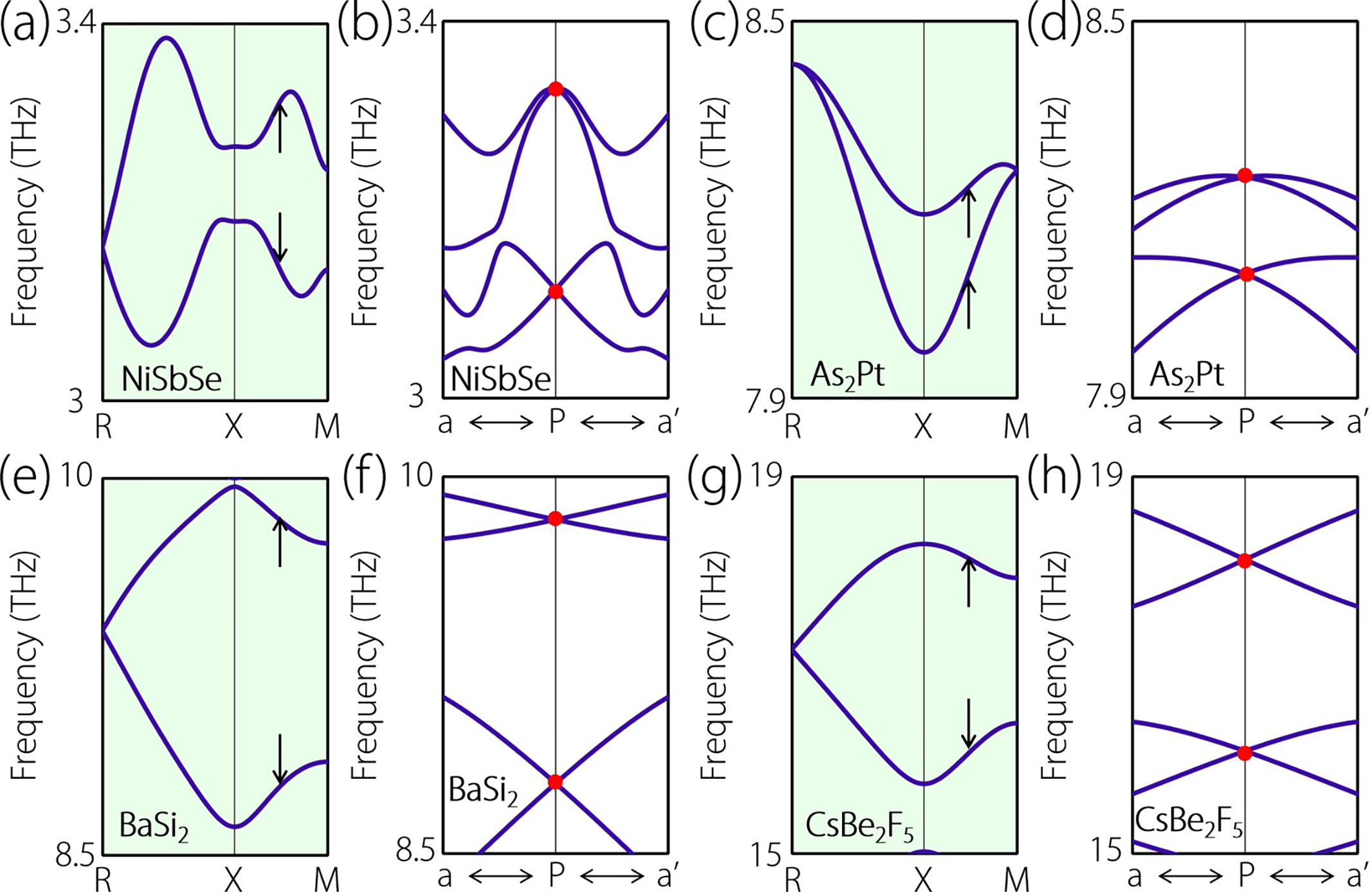}
\caption{(a), (c), (e), (g) Enlarged phonon dispersions of $R$-$X$-$M$ path (see Fig.~\ref{fig8}(b)--(e)) of As$_{2}$Pt, BaSi$_{2}$, and CsBe$_{2}$F$_{5}$ materials, respectively. (b), (d), (f), (h) Phonon dispersions along $a$ (0.00, 0.70, 0.25) -$P$ (0.00, 0.50, 0.25) -$a^{\prime}$ (0.00, 0.30, 0.25) path. All points at the $P$ symmetry point are two-fold degenerate points with linear phonon band dispersions (indicated by red circles in (b), (d), (f), and (h)).
\label{fig9}}
\end{figure}

We calculated the phonon dispersions of NiSbSe, As$_{2}$Pt, BaSi$_{2}$, and CsBe$_{2}$F$_{5}$ along the symmetry paths $\Gamma - X - M - \Gamma - R - X - M$ (see Fig.~\ref{fig8}(a)); the results are shown in Fig.~\ref{fig8}(b)--(d). Let us focus on the two-fold degenerate phonon bands along the $R$-$X$-$M$ paths (see Fig.~\ref{fig9}(a), (c), (e), and (g)). To prove that these bands are degenerated, we chose the path $a$-$P$-$a^{\prime}$ that vertically passes through the $k_{y}$ = $\pi$ plane. The obtained phonon dispersions along the $a$-$P$-$a^{\prime}$ path for these materials are shown in Fig.~\ref{fig9}(b), (d), (f), and (h), respectively. There are two evident two-fold degenerate points at $P$ with linear band dispersions. We can conclude that an NS phonon exists on the $k_{y}$ = $\pi$ plane on which the two low-energy phonon bands cross linearly. Owing to $C_{3,111}$ symmetry, equivalent NS phonons can be found on the $k_{x}$ = $\pi$ and $k_{y}$ = $\pi$ planes. We would like to point out that although symmetry requires the occurrence of three-NS phonons and limits the possible positions on the $k_{i}$ = $\pi$ ($i$ = $x$, $y$, $z$) planes, it does not limit the frequencies and dispersions of three-NS phonons.

\section{Summary and remarks}

In conclusion, according to the symmetry analysis results, there are three-NS phonons in the SGs with SG Nos. 19, 61, 62, 92, 96, 198, 205, 212, and 213 of the 230 SGs. More interestingly, by performing first-principles calculations, we discovered that the realistic materials $P2_{1}$2$_{1}$2$_{1}$-type YCuS$_{2}$ (SG No. 19), $Pbca$-type NiAs$_{2}$ (SG No. 61), $Pnma$-type SrZrO$_{2}$ (SG No. 62), $P4_{1}$2$_{1}$2-type LiAlO$_{2}$ (SG No. 92), $P4_{3}$2$_{1}$2-type ZnP$_{2}$ (SG No. 96), $P2_{1}$3-type NiSbSe (SG No. 198), $Pa\bar{3}$-type As$_{2}$Pt (SG No. 205), $P4_{3}$32-type BaSi$_{2}$ (SG No. 212), and $P4_{1}$32-type CsBe$_{2}$F$_{5}$ (SG No. 213) include three-NS phonons in their phonon dispersions.

We present the following remarks: (i) Because phonons obey Bose--Einstein statistics and are not limited by the Fermi energy, three-NS in the phonon system may be more common in realistic materials; (ii) Unlike fermions in electronic systems with heavy elements, SOC can be neglected for TPs in phonon systems. Hence, three-NS phonons in phonon systems can be considered real NS states without SOC-induced gaps; (iii) Although three-NS phonons in SGs 19, 61, 62, 92, 96, 198, 205, 212, and 213 can be determined by the combination of two-fold screw symmetry and time reversal symmetry, the frequencies and dispersions of three-NS phonons are not limited; (iv) One may ask what is the difference between three-NS and one-/two-NS states? As we know, constrained by the no-go theorem, among all the Weyl semimetals~\cite{add77} discovered in experiments before 2019, Weyl points always occur in pairs in the momentum space, without exception. Interestingly, in 2019, as demonstrated by Yu \textit{et al}.~\cite{add78}, the three-NS state is a good platform for realizing a singular Weyl point by circumventing the no-go theorem. However, for two- and one-NS states, although Weyl points and NS states can coexist, there must be more than one Weyl point in the BZ. Fortunately, in this year, Ma \textit{et al}.~\cite{add77} observed a singular Weyl point surrounded by three-NSs in PtGa with SG No. 198 in an experiment.

{\color{red}\textit{Acknowledgments}} X.T.W. thanks Prof. Zhi-Ming Yu for his help regarding to this manuscript. Y.L. is grateful for the support from the Nature Science Foundation of Hebei Province (No. A2021202002). X.T.W. is grateful for the support from the National Natural Science Foundation of China (No. 51801163) and the Natural Science Foundation of Chongqing (No. cstc2018jcyjA0765).

\end{document}